\documentclass[preprint, number,12pt]{elsarticle}

 \usepackage{graphics}
 \usepackage{graphicx}

\graphicspath{{figures//}}

\usepackage{amssymb}
\usepackage{amsthm}

\usepackage[nodots]{numcompress}

\DeclareGraphicsExtensions{.eps, .pdf,.png,.jpg,.tif}

\journal{Journal of Computational Science}

\begin{document}

\begin{frontmatter}

\title{Distributing points uniformly on the unit sphere under a mirror reflection symmetry constraint}

\author[aff01]{Cheng Guan Koay\corref{cor1}}  %%\fnref{fn1}}
\ead{cgkoay@wisc.edu}

%%\author[aff01,aff02]{Cheng Guan Koay\corref{cor1}}  %%\fnref{fn1}}
%%\ead{cgkoay@wisc.edu}

\cortext[cor1]{Corresponding author}
%\fntext[fn1]{Now at the Department of Medical Physics, University of Wisconsin-Madison, Wisconsin.}

\address[aff01]{
Department of Medical Physics\\
University of Wisconsin School of Medicine and Public Health\\
1161 Wisconsin Institutes for Medical Research (WIMR)\\
1111 Highland Avenue\\
Madison, WI 53705\\
}

%%\address[aff02]{Section on Tissue Biophysics and Biomimetics\\
%%Eunice Kennedy Shriver National Institute of Child Health and Human Development\\
%%National Institutes of Health\\
%%Bldg. 13, Rm. 3W16\\
%%13 South Drive, MSC 5772\\
%%Bethesda, MD 20892\\
%%}

%%

\begin{abstract}
Uniformly distributed point sets on the unit sphere with and without symmetry constraints have been found useful in many scientific and engineering applications. Here, a novel variant of the Thomson problem is proposed and formulated as an unconstrained optimization problem. While the goal of the Thomson problem is to find the minimum energy configuration of $N$ electrons constrained on the surface of the unit sphere, this novel variant imposes a new symmetry constraint---mirror reflection symmetry with the $x$-$y$ plane as the plane of symmetry. Qualitative features of the two-dimensional projection of the optimal configurations are briefly mentioned and compared to the ground-state configurations of the two dimensional system of charged particles laterally confined by a parabolic potential well.
\end{abstract}


\end{frontmatter}

%\mainmatter
\section{Introduction}\label{intro}
The essence of the Thomson problem \cite{Thomson1904,Fekete1923} is to find the minimum electrostatic potential energy configuration of $N$ electrons constrained on the surface of the unit sphere. Even though the Thomson problem has been around for more than a century now, it never ceases to inspire new developments and applications. From the aesthetic point view \cite{Smale1998}, the Thomson problem rivals some of the famous unsolved problems in number theory in terms of its aesthetic appeal such as the simplicity of the problem statement, the complexity of the general solution, the computability or tractability of some simple cases and the beauty of the minimum-energy configurations. More importantly, it is its immense practical appeal that will make it relevant well into the future. Specifically, the minimum energy configurations obtained from the Thomson problem or other uniform point sets on the unit sphere through other approaches have been found useful, and in some cases essential, in many scientific and engineering applications \cite{Basser1994, Jones1999a, Bauer00, Peters2000, Barger2002, Koay2006, Wu2007, Mitchell2008,  Koay2011a, Koay2011b,Koay2011c, King2012, Koay2012, Caruyer2013, Koay2014}.

Among the many developments, we list the following research directions that have been inspired in part by the Thomson problem:
\begin{itemize}
\item The development of a deterministic scheme capable of generating nearly uniform points on the unit sphere, see \cite{Saff94,Saff97,Bauer00,WongMRM94,Koay2011a}.

\item The development of a new variant of the Thomson problem under the antipodal symmetry constraint. The approach that deals directly with the repulsive forces was first suggested in \cite{Jones1999a} and elaborated with further details in \cite{De Santis2014}. The reformulation of the electrostatic potential energy function to account for antipodal symmetry can be found in \cite{Koay2012, Koay2014}. A new iterative scheme based on the centroidal Voronoi tessellations \cite{Du1999} capable of generating large-scale uniform antipodally symmetric points on the unit sphere suitable for 3D radial MRI applications was proposed and developed in \cite{Koay2014}. In  \cite{Koay2014}, a novel pseudometric was proposed and derived from the modified Coulomb interaction term studied in \cite{Koay2012}.

\item Development of a deterministic scheme capable of generating nearly uniform and antipodally symmetric points on the unit sphere \cite{Koay2011b, Koay2011c}. This scheme is based on a set of equidistant latitudes under the condition that the spacing between two consecutive points on the same latitude should be approximately equal to the spacing between two consecutive latitudes for a given number of points. This type of point set has been found useful in the analysis of astrophysical data \cite{King2012}, medical imaging \cite{Chang2014} and in numerical works in optical sciences\cite{Ranasinghesagara14} \footnote{The author learned of the application of this type of point set in optical sciences through a private correspondence with Dr. Janaka Chamith Ranasinghesagara of University of California, Irvine.}.
\end{itemize}

In this communication, we propose a novel variant of the Thomson problem under a mirror reflection symmetry constraint. The reformulation of the electrostatic potential energy function to account for the mirror reflection symmetry is similar to that of our previous work in dealing the Thomson problem under the antipodal symmetry constraint, \cite{Koay2012,Koay2014}. We will also reformulate the Coulomb interaction term so that a new pseudometric can be gleaned from it and can be used in a similar framework to that of our pseudometrically constrained centroidal Voronoi tessellations of the unit sphere \cite{Koay2014}.

Even though our interest lies mainly on generating uniformly distributed points on the upper hemisphere for modeling and analysis purposes. The two-dimensional projections of these point sets on the $x$-$y$ plane turns out to have apparent similarity to the arrangement of charged particles in a two-dimensional system confined by a potential well, which has been found useful in many atomic and condensed matter physics applications, see \cite{Lozovik90,Bolton93,Bedanov94,SaintJean01,Reimann02}. We should point out that the optimal configurations obtained from the proposed problem is well suited for applications in optics, especially in the modeling of spherical lenses because the boundary of the proposed uniform point sets is the equatorial great circle. Consequently, it is a simple task to map those uniformly distributed points on the upper hemisphere onto a spherical cap to facilitate the utility of this type of point set in optical and other imaging sciences.  Another potentially interesting and useful application of the proposed point set that is closer to the author's interest is in making this kind of point sets as a coordinate system for mapping brain (hemispheric or left-right) asymmetry \cite{Toga2003,Chung2001}.

\section{Methods}\label{method}
The spherical coordinate on the unit sphere can be described by, $(\theta,\phi)$, and its transformation to Cartesian coordinate, $(x,y,z)$, can be accomplished by the following transformations:
\begin{eqnarray*}
x &=& \sin(\theta) \cos(\phi), \\
y &=& \sin(\theta) \sin(\phi), \\
z &=& \cos(\theta).
\end{eqnarray*}
The mirror reflection operation about the $x$-$y$ plane is denoted by $\sigma:\mathbb{R}^{3}\rightarrow \mathbb{R}^{3}$. For example, it maps a vector, $\mathbf{r}=[r_{x},r_{y},r_{z}]^{T}$, on the upper unit hemisphere to its mirror image, $\mathbf{x}=\sigma(\mathbf{r}) = [r_{x},r_{y}, -r_{z}]^{T}$ and vice versa. Note that the vector or matrix transposition is denoted by $T$. The matrix representation of $\sigma$ is a diagonal matrix with $\{1, 1, -1\}$ in the main diagonal.

Suppose we have a collection of $2N$ points, $\{\mathbf{r}_{1},\cdots,\mathbf{r}_{N}, \mathbf{x}_{1},\cdots,\mathbf{x}_{N}\}$, on the unit sphere and that this collection of points is endowed with mirror reflection symmetry, i.e., $\mathbf{x}_{i}=\sigma(\mathbf{r}_{i})$ for $i=1,\cdots, N$. We further assume that $\mathbf{r}_{i}'s$ are on the upper hemisphere. In general, the electrostatic potential energy of $2N$ points, say $\{\mathbf{y}_{1},\cdots,\mathbf{y}_{2N}\}$, is typically expressed as a double summation of $2N$ terms
\begin{eqnarray*}
\sum^{2N}_{i=1}\sum^{2N}_{j=i+1} \frac{1}{||\mathbf{y}_{i}-\mathbf{y}_{j}||}.
\end{eqnarray*}
Due to the mirror reflection symmetry constraint, the electrostatic potential energy of $2N$ points can now be expressed as a combination of a double sum of $N$ terms and a single sum of $N$ terms
\begin{eqnarray}\label{EQ1}
\phi_{C} = 2 \sum^{N}_{i=1}\sum^{N}_{j=i+1} \left(\frac{1}{||\mathbf{r}_{i}-\mathbf{r}_{j}||} + \frac{1}{||\mathbf{r}_{i}-\sigma(\mathbf{r}_{j})||} \right) + \sum^{N}_{k=1} \frac{1}{||\mathbf{r}_{k}-\sigma(\mathbf{r}_{k})||}.
\end{eqnarray}
The first term and the second term in the double sum account for the interaction between any two distinct points on the upper hemisphere and between a point on the upper hemisphere and another point in the lower hemisphere that is not its mirror image, respectively. The term in the single sum accounts for the interaction between a point on the upper hemisphere and its mirror image. The above reformulation is similar to that of the Thomson problem under antipodal symmetry constraint \cite{Koay2012,Koay2014} except the appearance of the second sum in Eq.~(\ref{EQ1}). The appearance of this second sum makes it harder than the case under the antipodal symmetry constraint to extract the desired pseudometric, which can then be used in a framework similar to our previous proposed pseudometrically constrained centroidal Voronoi tessellations, see Eqs.(3-4) in \cite{Koay2014}, to deal with large-scale problems, e.g., generating large number of uniformly distributed points in the range of tens of thousands. In order to extract the desired pseudometric, we would like the interaction term to appear within the double sum. It is not hard to see that the following expression is equivalent to Eq.~(\ref{EQ1}):
\begin{eqnarray}\label{EQ2}
\phi_{C} &=& \sum^{N}_{i=1}\sum^{N}_{
  j=1,j\neq i
} \left( \frac{1}{||\mathbf{r}_{i}-\mathbf{r}_{j}||} + \frac{1}{||\mathbf{r}_{i}-\sigma(\mathbf{r}_{j})||} \right.\nonumber\\
     & & \left. + \frac{1}{2(N-1)}\left[ \frac{1}{||\mathbf{r}_{i}-\sigma(\mathbf{r}_{i})||}  + \frac{1}{||\mathbf{r}_{j}-\sigma(\mathbf{r}_{j})||}  \right]\right).
\end{eqnarray}
The desired pseudometric, denoted by $d(\mathbf{r}_{i},\mathbf{r}_{j})$, is simply the reciprocal of the collective summand in Eq.~(\ref{EQ2}), i.e.,
\begin{eqnarray*}
d(\mathbf{r}_{i},\mathbf{r}_{j}) = \frac{1}{\frac{1}{||\mathbf{r}_{i}-\mathbf{r}_{j}||} + \frac{1}{||\mathbf{r}_{i}-\sigma(\mathbf{r}_{j})||} + \frac{1}{2(N-1)}\left[ \frac{1}{||\mathbf{r}_{i}-\sigma(\mathbf{r}_{i})||}  + \frac{1}{||\mathbf{r}_{j}-\sigma(\mathbf{r}_{j})||}  \right]}.
\end{eqnarray*}

\subsection{Qualitative relationship to the 2D system of charged particles in a circular parabolic potential well.}
The potential energy of two-dimensional system of $N$ classical charged particles confined within a circular parabolic potential well \cite{Lozovik90,Bolton93,Bedanov94} is given by the following expression
\begin{eqnarray}\label{EQ3}
 E=   \sum^{N}_{i=1}\sum^{N}_{j=i+1} \frac{1}{||\mathbf{r}_{i}-\mathbf{r}_{j}||} + \sum^{N}_{k=1} ||\mathbf{r}_{k}||^{2},
\end{eqnarray}
where $\mathbf{r}_{i}$'s are in $\mathbb{R}^{2}$.

Qualitatively, the effect of the second sum in Eq.(\ref{EQ3}), which comes from the the circular parabolic potential well, is to attract charged particles closer to the center of the potential well. Similarly, the effect of the second sum in Eq.(\ref{EQ1}), which comes from the mirror-symmetry constraint, is to repel charged particles that are confined to the surface of the sphere from the $x-y$ plane, which has the same effect of pushing charged particles closer to the z-axis. Due to the qualitative effects of the potential well and the plane of symmetry on charged particles, it is not surprising that the two-dimensional projections of the optimal configurations of the proposed problem may share some qualitative features to those of a system of charged particles in the circular parabolic potential well. Shell-like structures are common features in both systems.

In the limit when the upper hemisphere reduces to a very small spherical cap centered on the z-axis, the interactions between points on the upper hemisphere become more dominant than the inter-hemispheric interactions. Further, the first non-constant and dominant term in the series expansion of the term in the second sum in Eq.(\ref{EQ1}) goes as $r^{2}$ where $r$ is the planar distant between the position of the vertical projection of the point on the $x-y$ plane and the center of the $x-y$ plane. Due to the $r^{2}$ dependence, this limiting case is the closest in spirit to the two-dimensional system of charged particles in the circular parabolic potential well.

\section{Results}\label{Results}
Any nonlinear optimization solver requires an initial solution to find the optimal solution and it is well known that the solver reaches the optimal solution more readily with lower likelihood of getting stuck in local minima if the initial solution is  \textit{close} to the optimal solution. It is interesting to note that the point sets generated by our previous deterministic scheme, which was designed to generate nearly uniform and antipodally symmetric points on the upper unit hemisphere \cite{Koay2011b, Koay2011c}, can also be used as an initial solution for our current variant of the Thomson problem under mirror-reflection symmetry constraint. As will be discussed later, empirical results show that the point sets generated by our deterministic scheme turns out be very well suited for the optimization problem considered here.

We implemented the proposed optimization problem in Java on a machine with Four $\times$ 8-Core 2.3GHz AMD Opteron Processors (6134) with 64 GB of RAM. We used the Broyden--Fletcher--Goldfarb--Shanno (BFGS) algorithm \cite{Dennis1996} to solve the proposed optimization problem for a wide range of $N$, i.e., from $N=2$ to $N=500$. The analytical expressions of the gradient of the objective function (Eq.~(\ref{EQ1})) with respect to the parameters of interest (in spherical coordinates) were used within the BFGS algorithm. These expressions are listed in the Appendix. The inverse of the Hessian matrix was constructed approximately at the first iteration and updated from one iteration to the next. The convergence criterion adopted here is based on the norm of the gradient of the objective function. The solution is deemed optimal if the norm of the gradient of the objective function evaluated at the solution is less than some fixed tolerance ( $1.0\times10^{-8}$ was chosen as the tolerance level in this study).

We would like to point out one important detail about our implementation. Even though we are dealing with points on the upper hemisphere ( $0\leq\theta_{i}\leq\pi/2$ and $0\leq\phi_{i}\leq2\pi$ for all $i$), it may happen that some $\theta_{i}$ may get out of bound during the update step within the BFGS algorithm. The strategy adopted here is to map the affected point back to its mirror image in the upper hemisphere, i.e., $\theta_{i}=\pi-\theta_{i}$ if $\theta_{i}>\pi/2$.

Of the 499 cases ($N=2,\cdots,500$) tested in this study, only four cases ($N=124,155,341,444$) did not converge using the initial solutions obtained from \cite{Koay2011b}. However, a small and random perturbation can be applied to these initial solutions to help the optimization solver reach convergence. We investigated the case of $N=444$ in more detail and found that the convergence could be reached by invoking the BFGS algorithm again using the non-convergent point set generated by the first invocation of the BFGS algorithm after the maximum number of iterations was reached . This finding points to the possibility that the approximate inverse of the Hessian constructed and updated by the BFGS algorithm might have been suboptimal. Therefore, it may be advisable to re-initialize the BFGS algorithm should the maximum number of iterations is reached. It is also clear from these empirical results that the point sets generated by our deterministic scheme were very well suited for the current optimization problem.

\begin{figure}[htb]
\begin{center}
\leavevmode
\includegraphics[width=7.0cm]{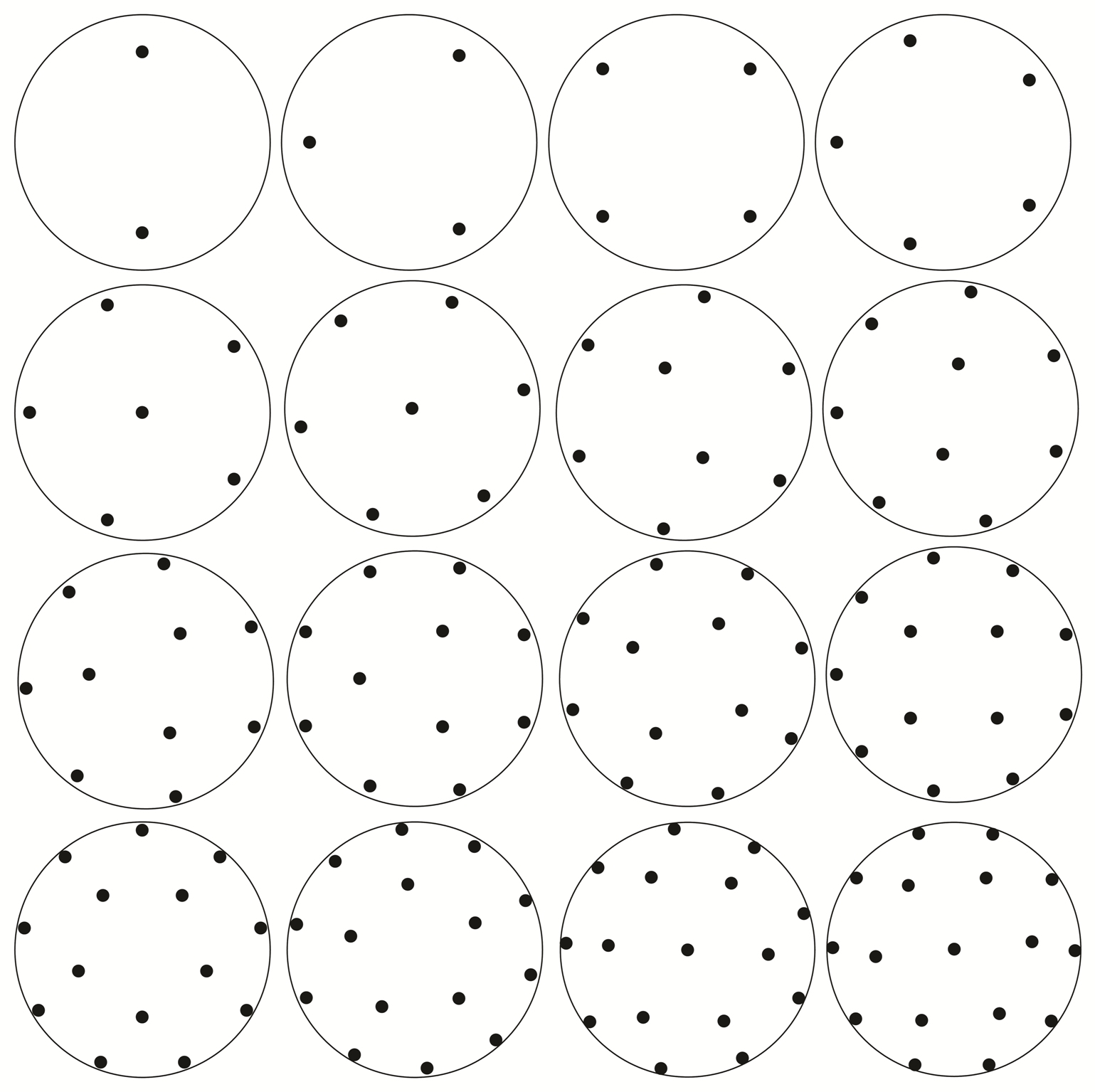}
\end{center}
\caption{The projection of the optimal configuration of the Thomson problem under mirror-reflection mirror constraint on the $x$-$y$ plane for $N=2,\cdots,17$.}
\label{F1}
\end{figure}

\begin{figure}[htb]
\begin{center}
\leavevmode
\includegraphics[width=8.0cm]{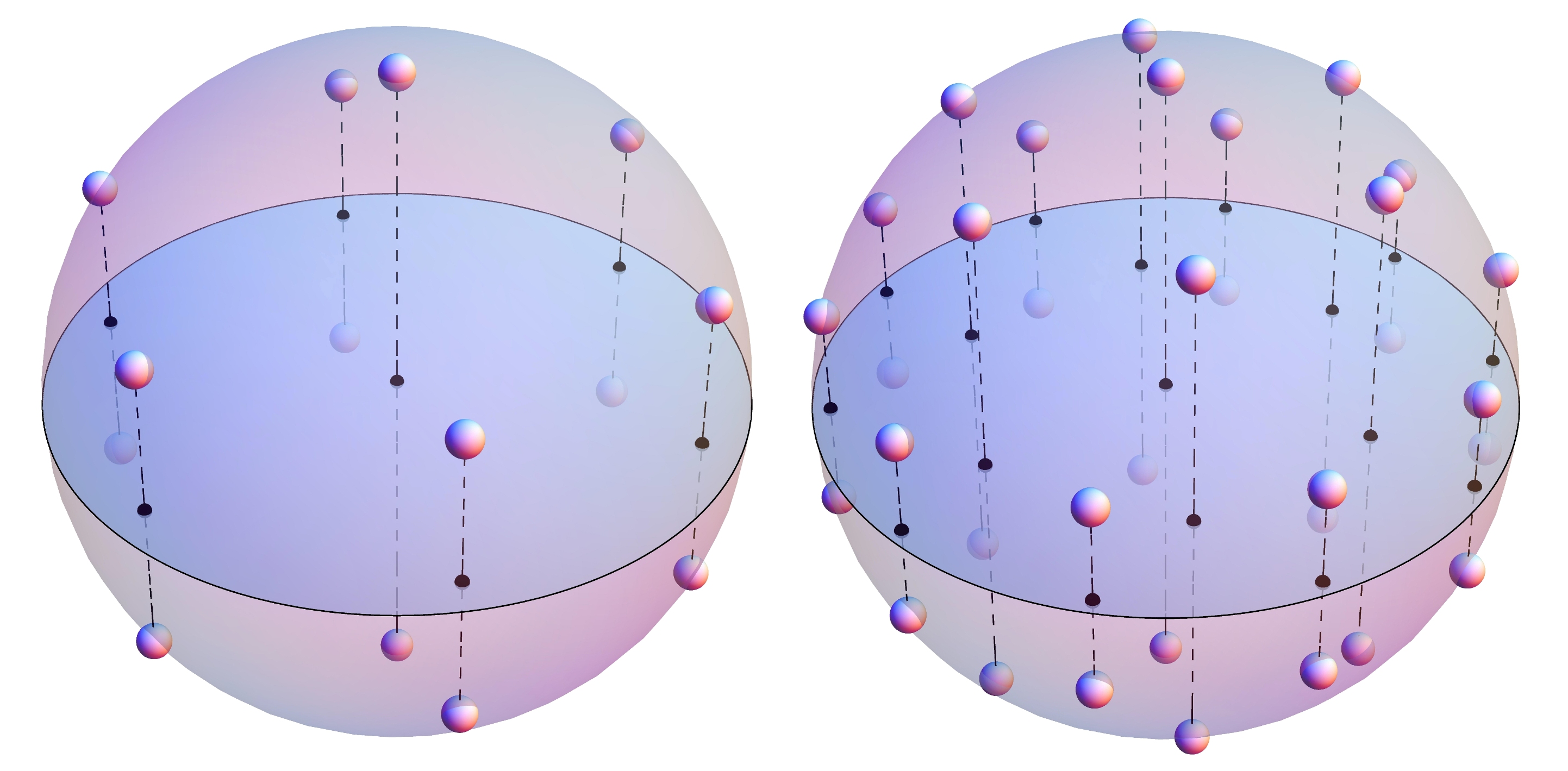}
\end{center}
\caption{Three-dimensional view of the optimal configuration for $N=7$ and $N=17$.}
\label{F2}
\end{figure}

It is interesting to note that a unique solution exists for the case of $N=1$. The unit vector along the $z$-axis together with its mirror-image on the negative $z$-axis is the optimal configuration. The optimal configuration for any $N$ greater than unity is unique up to rotation about the z-axis. The two-dimensional projection of the first 17 optimal configurations excluding the case of $N=1$ on the $x$-$y$ plane are shown in Fig.~\ref{F1}. The three-dimensional view of two of the optimal configurations, $N=7$ and $N=17$, are shown in Fig.~\ref{F2}. The fact that every optimal configuration displayed here in Fig.~\ref{F1} contains nearly symmetrical polygonal patterns of distinct radii is very intriguing because similar `shell-like' structures have been found in two-dimensional systems of charged particles under confinement \cite{Lozovik90,Bolton93,Bedanov94}; some of the configurations ($N=2,\cdots,N=6$) seem to have additional discrete rotational symmetries ($n$-fold rotation about the $z$-axis with $n=1,\cdots,6$). A quick comparison of the two-dimensional optimal configurations of the proposed problem to those of the two-dimensional charged particles in a circular parabolic potential well \cite{Bedanov94} shows that the configurations with the following numbers of points, $1,\cdots, 7, 9, 13, 15, 17$, are equivalent at least topologically. No equivalent configuration was found beyond $N=17$. The proposed configurations seem to be more densely packed near the center of the $x-y$ plane and less densely packed near the boundary, e.g., the $N=14$ system has 5 and 9 points on the inner and outer shells, respectively, than the two-dimensional system under a parabolic potential well, i.e., the two-dimensional system of 14 points has 4 and 10 points on the inner and outer shells, respectively. The present system makes the transition from a 3-shell structure to a 4-shell structure at $N=28$ whereas the system under the circular parabolic potential well make the transition at $N=32$. Interestingly, both systems make the first two transitions at the same $N$, i.e., $N=6$ and $N=16$.

\begin{figure}[htb]
\begin{center}
\leavevmode
\includegraphics[width=8.0cm]{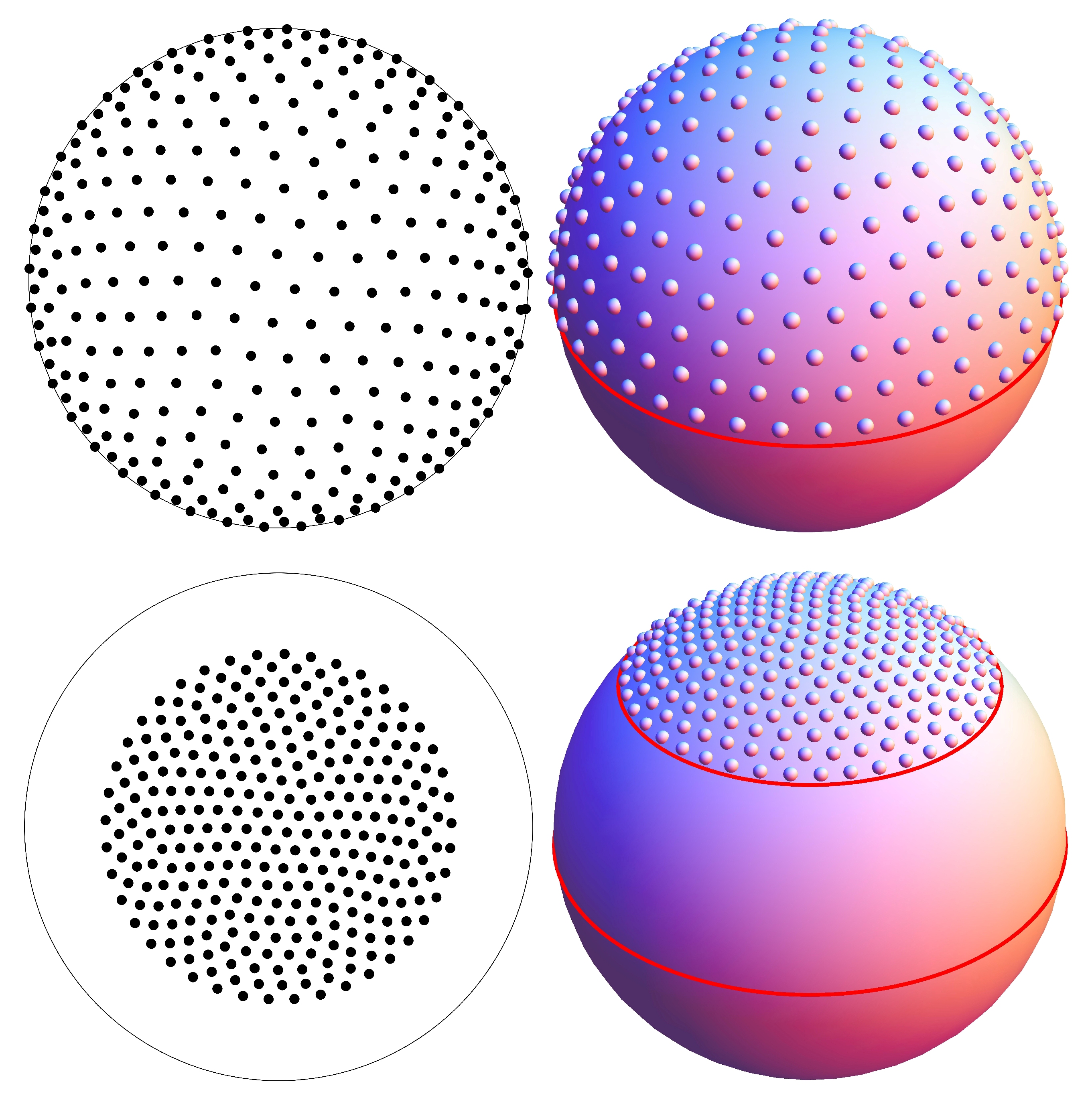}
\end{center}
\caption{Two-dimensional and three-dimensional view of the proposed point set with N=288 on the upper hemisphere and the corresponding transformed point set on a spherical cap with a maximum $\theta$ of $\pi/4$.}
\label{F3}
\end{figure}

As mentioned in the introduction, it is a simple task to map the proposed point set on the upper hemisphere onto a spherical cap with a maximum $\theta$ of $\alpha$, i.e., each point is transformed according to the following mapping, $(\theta,\phi)\longmapsto (\theta\frac{\alpha}{\pi/2},\phi)$ in spherical coordinates, see Fig.~\ref{F3} for an example with $N=288$ on the upper hemisphere.

\section{Discussion}\label{Discussion}
The goal of this brief communication is to share with interested readers a new approach of distributing points uniformly on the sphere that are constrained by a new symmetry---mirror reflection symmetry. This work and our previous work \cite{Koay2012,Koay2014} show a general strategy of reformulating the electrostatic potential energy function to account for different symmetries (the mirror reflection and the antipodal symmetries), respectively. It should be clear that such a strategy can be adapted to solving other variants of the Thomson problem under more specialized symmetries such as the tetrahedral symmetry or higher-order symmetries.

The unique feature of the proposed optimal configurations is that every optimal configuration shares the same boundary, which is the great circle at the equator. Based on our previous experience with the Thomson problem under the antipodal symmetry constraint \cite{Koay2012,Koay2014}, it can be concluded that the optimal configurations (with antipodal symmetry) do not share the same boundary, see Fig.~3B in \cite{Koay2014}. This unique feature of the proposed configurations is likely to be useful in applications where a well-defined boundary is required. Therefore, the present work is likely to be found useful in many imaging sciences and geosciences. Finally, interested readers are invited to explore diverse applications in which new and novel constraints are placed on uniformly distributed points on the sphere to enhance imaging acquisitions in Magnetic Resonance Imaging \cite{Koay2011c,Koay2012,Caruyer2013,De Santis2014, Koay2014}.

The limitation of any unconstrained optimization algorithm of multivariable functions is that the algorithm usually does not scale well with the size of the unknown parameters. This is particularly true with the BFGS algorithm used because the approximate inverse of the Hessian matrix is constructed iteratively and the $O(N^2)$ complexity associated with the evaluation of the potential energy function. The limited memory version of BFGS \cite{Nocedal1980}, which keeps a few dominant vectors to approximate the inverse of the Hessian matrix, may be useful for solving the proposed problem for $N > 500$. However, it is unlikely that the limited memory version of BFGS will be effective when the number of unknown parameters is more than several thousands.

Due to the similar qualitative features of the potential energy function of the proposed problem and that of a two-dimensional system of charged particles in a circular parabolic potential well, we believe the two-dimensional projection of the point set generated by our previously proposed deterministic method \cite{Koay2011b, Koay2011c}, can be used as a quick and efficient initial solution to any work simulating a two-dimensional system of charged particles in a circular parabolic potential well.

\section*{Acknowledgments}
The author dedicates this work to the memory of Dr. Amer Lahamer. The author would like to express his sincere thanks to Dr. Peter J. Basser and Prof. Beth Meyerand for their continued support. He would also like to thank Dr. Janaka Chamith Ranasinghesagara of University of California---Irvine for helpful discussion. He also would like to thank one of the anonymous reviewers for pointing out the apparent similarity of the two-dimensional projections of the proposed optimal configurations to those works involving Wigner lattice, two-dimensional systems of charged particles or quantum dots that exhibit electron shells in a Mendeleev-like table. Software related to this work will be made available at http://sites.google.com/site/hispeedpackets/.

\section*{Appendix A}
In this appendix, we list the expressions for the derivative of the electrostatic potential energy function with respect to the spherical coordinates of each of the points on the upper hemisphere. For convenience, we define $s_{ij} = ||\mathbf{r}_{i}-\mathbf{r}_{j}||$ and $\tilde{s}_{ij} = ||\mathbf{r}_{i}-\sigma(\mathbf{r}_{j})||$. The derivative of $\phi_{C}$ with respect to $\theta_{i}$ (the first component of the spherical coordinates of the $i$-th point on the upper hemisphere) is given by
\begin{eqnarray*}
\frac{\partial\phi_{C}}{\partial \theta_{i}} = - 2 \sum^{N}_{j\neq i} \left(\frac{1}{s^{2}_{ji}} \frac{\partial s_{ji}}{\partial \theta_{i}} + \frac{1}{\tilde{s}^{2}_{ji}} \frac{\partial \tilde{s}_{ji}}{\partial \theta_{i}} \right) + \frac{\sin(\theta_{i})}{2\cos^{2}(\theta_{i})},
\end{eqnarray*} and the derivative of $\phi_{C}$ with respect to $\phi_{i}$ is given by
\begin{eqnarray*}
\frac{\partial\phi_{C}}{\partial \phi_{i}} = - 2 \sum^{N}_{j\neq i} \left(\frac{1}{s^{2}_{ji}} \frac{\partial s_{ji}}{\partial \phi_{i}} + \frac{1}{\tilde{s}^{2}_{ji}} \frac{\partial \tilde{s}_{ji}}{\partial \phi_{i}} \right).
\end{eqnarray*}
%
%
%\begin{figure}[htb]
%\begin{center}
%\leavevmode
%\includegraphics[width=7.0cm]{Figure4.tif}
%\end{center}
%\caption{Modified spherical cap for selecting neighboring generators when the generator is closest to the equatorial great circle.}
%\label{F4}
%\end{figure}

\bibliographystyle{elsarticle-harv}
%%\bibliography{MyDatabase}

\end{document}